\documentclass[epj]{svjour}
\usepackage{bm,graphicx,amsmath}

\begin{document}
\title{Archimedean-like colloidal tilings on substrates with decagonal and tetradecagonal symmetry}

\author{Michael Schmiedeberg\inst{1}\and Jules Mikhael\inst{2} \and 
Sebastian Rausch\inst{2} \and Johannes Roth\inst{3}\and 
Laurent Helden\inst{2} \and Clemens Bechinger\inst{2,4}\and 
Holger Stark\inst{5}}
\institute{Department of Physics and Astronomy, University of Pennsylvania, 
209 South 33rd Street, Philadelphia, PA 19104, USA
\and
2. Physikalisches Institut,  
Universit\"at Stuttgart, Pfaffenwaldring 57, D-70550 Stuttgart, Germany
\and 
Institut f\"ur Theoretische und Angewandte Physik, 
Universit\"at Stuttgart, Pfaffenwaldring 57, D-70550 Stuttgart, Germany
\and 
Max-Planck-Institut f\"ur Metallforschung, Heisenbergstra\ss e 3, 
D-70569 Stuttgart, Germany
\and 
Institut f\"ur Theoretische Physik, Technische Universit\"at 
Berlin, Hardenbergstr. 36, D-10623 Berlin, Germany}
\authorrunning{M. Schmiedeberg \emph{et al.}}
\titlerunning{Archimedean-like colloidal tilings on substrates with decagonal 
and tetradecagonal symmetry}
\date{}
\abstract{
Two-dimensional colloidal suspensions subject to laser interference 
patterns with decagonal symmetry can form an Archimedean-like tiling 
phase where rows of squares and triangles order aperiodically along one
direction [J. Mikhael \emph{et al.}, Nature {\bf{454}}, 501 (2008)].
In experiments as well as in Monte-Carlo and Brownian dynamics simulations,
we identify a similar phase when the laser field possesses tetradecagonal 
symmetry  We characterize the structure of both Archimedean-like 
tilings in detail and point out how the tilings differ from each other.
Furthermore, we also estimate specific particle densities where the 
Archimedean-like tiling phases occur. Finally, using Brownian dynamics 
simulations we demonstrate how phasonic distortions of the decagonal
laser field influence the Archimedean-like tiling. In particular,
the domain size of the tiling can be enlarged by phasonic drifts and
constant gradients in the phasonic displacement. We demonstrate 
that the latter occurs when the interfering laser beams are not adjusted
properly.
\PACS{
{61.44.Br}{Quasicrystals}\and
{82.70.Dd}{Colloids}
}
}
\maketitle


\section{Introduction} \label{intro}

\begin{figure}
\begin{center}
  \includegraphics[width=0.6\linewidth]{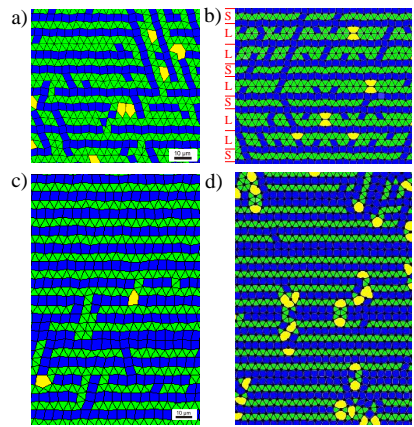}
\end{center}
  \caption{Archimedean-like tiling phases 
induced by (a,b) the decagonal and (c,d) the tetradecagonal substrate. 
(a,c) show experimental results, (b,d) are 
obtained by Monte-Carlo simulations. The figures show Delaunay triangulations 
of colloidal point patterns. 
Bonds longer than $1.1a_S$  ($1.2a_S$ for experiments) are omitted, where 
$a_{S}$ is the spacing of the particles 
when placed on an ideal triangular lattice. 
The potential strength is $V_0/(k_BT)=13$ in (a), $V_0/(k_BT)=10$ in (c), and $V_0/(k_BT)=20$ in the simulations (b,d). The density is given by $a_S/a_V=0.57$ in (a), $a_S/a_V=0.58$ in (b), and $a_S/a_V=0.7$ in (c,d).}
\label{fig:AT-sim} 
\end{figure}

Unlike ordinary crystals, quasicrystals possess point-sym\-me\-tries, e.g., 
with five- or ten-fold rotational axes, which are not allowed in periodic 
crystals \cite{shechtman84,levine84}. Nevertheless, quasicrystals 
exhibit Bragg-reflections due to their long-range positional and 
orientational order. 
Since they show exceptional material properties \cite{macia06}, there is
much interest in understanding and controlling the growth of quasicrystals.
Therefore, a lot of research 
has been devoted to 
study the ordering of atoms on quasicrystalline 
surfaces \cite{McGrath02a,McGrath02b,Shimoda00,Franke02,Ledieu04,Tasca04a,Tasca04b,Tasca04c,Tasca04d,Bilki07a,Bilki07b,Pussi09}. 
However, in these 
experiments it is very difficult to determine the exact positions 
of the atoms.
Therefore, a model system  consisting of micron-sized colloidal 
particles subject to a laser interference pattern
has been used to investigate the dynamics and ordering of particles in a 
two-dimensional quasicrystalline potential 
\cite{schmiedeberg06,schmiedeberg07,mikhael08,schmiedeberg08,mikhael09}. 
Interesting structures have been observed, such as phases with 20 bond 
directions \cite{schmiedeberg08} or colloidal orderings consisting of rows 
of squares and rows of triangles termed Archimedean-like tilings 
\cite{mikhael08,schmiedeberg08} (see also Fig.\ \ref{fig:AT-sim}). The 
latter will be the main topic of this work.

\begin{figure*}
\begin{center}
  \includegraphics[width=\linewidth]{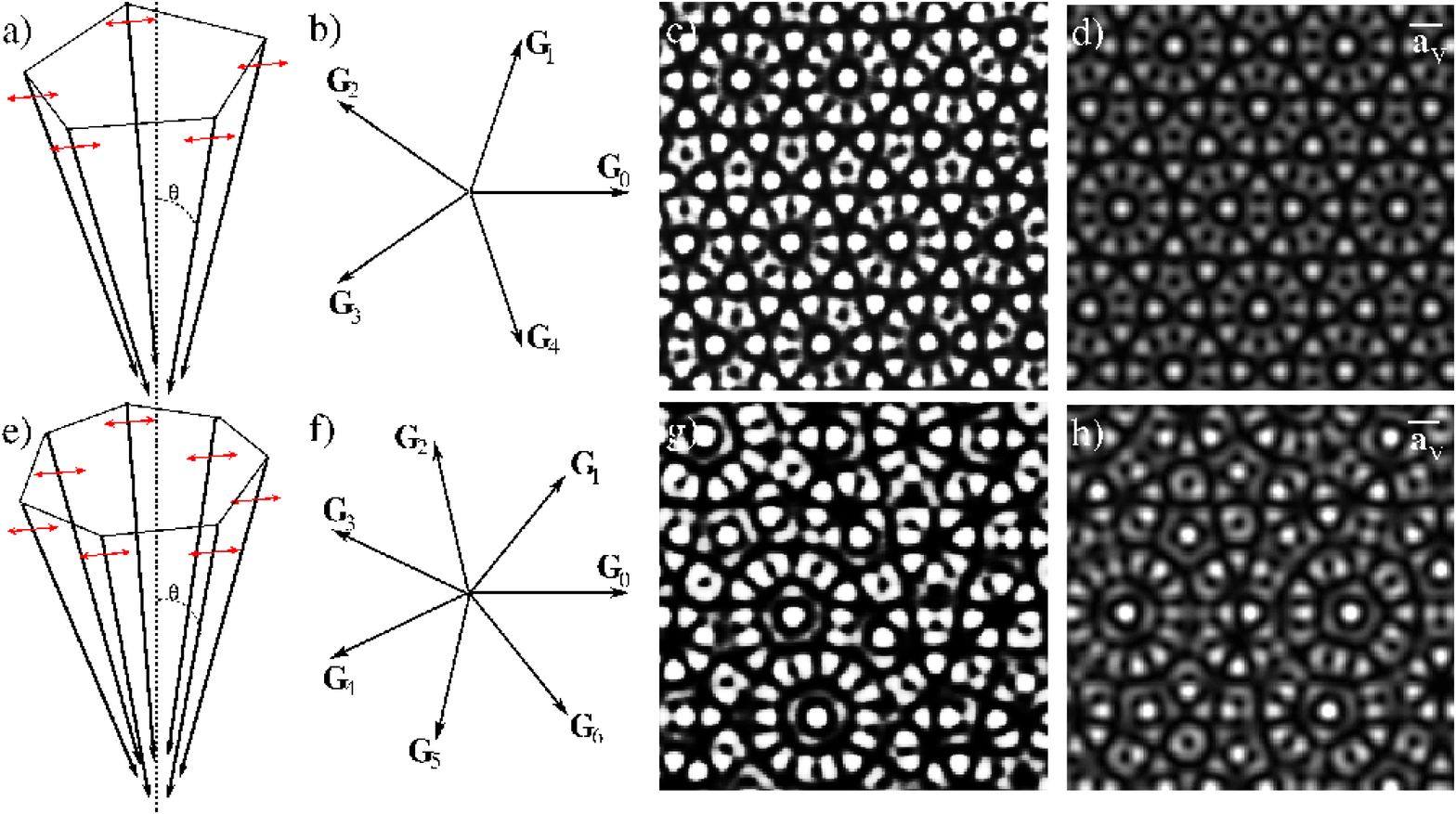}
\end{center}
  \caption{The external potentials are created by interfering laser beams
with a symmetric arrangement.
An interference pattern
with 10-fold rotational symmetry is obtained with five laser 
beams (a-d), 14-fold symmetry is realized with seven beams (e-h). (a,e) 
show how the laser beams are arranged
and (b,f) the wave-vectors projected onto the sample plane.
(c,g) demonstrate the interference patterns observed in experiments, and 
in (d,h) they are calculated from Eq.\ (\ref{eq:pot}).
The bar in the upper right corners of (d,h) denote the length scale $a_V=2\pi/ 
\left|\mathbf{G}_j\right|$, where $\mathbf{G}_j$ are the projected wave 
vectors as shown in (b,f).}
  \label{fig:pot} 
\end{figure*}

\begin{figure}
\begin{center}
  \includegraphics[width=0.9\linewidth]{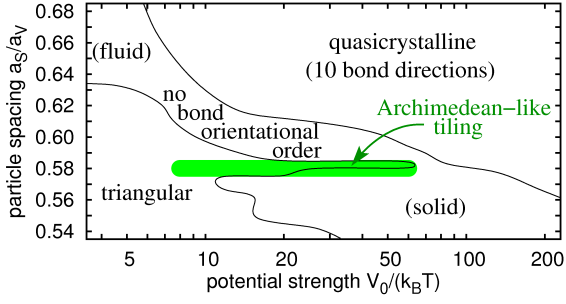}
\end{center}
  \caption{Phase diagram obtained with Monte-Carlo simulations for colloids 
on the decagonal substrate (the complete phase diagrams are presented and 
explained in \cite{schmiedeberg08}). Unlike quasicrystalline phases that we observed at all densities, Archimedean-like tiling structures are 
only found for intermediate potential strengths within a small range of 
densities (marked green).}
  \label{fig:pd-at} 
\end{figure}

Interference patterns of five or seven laser beams have decagonal or 
tetradecagonal symmetry, respectively (see Fig.\ \ref{fig:pot}). In such 
laser fields, the colloids are forced towards the regions with highest 
laser intensity, i.e., for the colloidal particles the interference 
patterns act like a substrate with quasicrystalline symmetry. If the 
influence of the laser field is weak, the colloidal particles form a 
triangular lattice in case of sufficiently strong repulsion forces or 
they are in a fluid phase 
when colloidal interactions are weak.
For large laser intensities, the ordering of the colloids is determined 
by the interference pattern, i.e., a quasicrystalline phase is observed 
(see also the phase diagrams in Fig.\ \ref{fig:pd-at} and in 
\cite{schmiedeberg08}). In this work we are especially interested in 
phases that occur at intermediate laser strengths where there is a 
competition between the repulsive colloidal interactions (that alone would 
lead to periodic triangular ordering) and the external forces due to the 
laser field which favor aperiodic quasicrystalline structures. For certain 
particle densities that we will determine later in this article, the 
competition leads to a colloidal ordering that consists of rows of triangles 
and rows of squares
that establish an aperiodic order in one spatial direction.
(see Fig.\ \ref{fig:AT-sim}). 
The structure observed is close to an Archimedean tiling, 
where the rows of squares and triangles form a two-dimensional periodic
order. It was therefore named Archimedean-like tiling when it was first 
realized in experiments using laser fields with decagonal 
symmetry \cite{mikhael08}. 
Later Archimedean-like tilings also occured in simulations that studied
the complete phase behavior on a decagonal substrate
(\cite{schmiedeberg08}, see also Fig.\ \ref{fig:pd-at}).

The purpose of this article 
is to characterize the structure of the 
Archimedean-like tiling phases in detail, to discuss the differences 
of the ordering in laser fields with decagonal and 
tetradecagonal symmetry, and finally
to determine the densities where these phases occur. 
In addition, we study the influence of phasonic distortions
on colloidal ordering in laser fields with decagonal symmetry. Phasons are 
excitations that only exist in quasicrystals. Like phonons, they are 
hydrodynamic modes, i.e., they do not cost free energy in the 
long-wave-length limit \cite{levine85a,levine85b}. 
Phasons are an ongoing main topic in current research on quasicrystals
and intensively discussed in the field \cite{henley06,freedman07a,freedman07b}.

In Sec.\ \ref{sec:model} we introduce our system. The details of the 
experimental realization as well as the numerical simulations 
are explained in Sec.\ \ref{sec:methods}. In Sec.\ \ref{sec:structure} we 
analyze the structure of the Archimedean-like tilings and calculate the 
densities where these phases occur. We study the consequences of phasonic 
distortions on the Archimedean-like tilings in Sec.\ \ref{sec:phasons} and finally conclude in Sec.\ \ref{sec:conclusions}.

\section{Model} 
\label{sec:model}

A charged stabilized suspension is confined between two glass plates and 
subjected to laser interference patterns
with quasicrystalline symmetries. Five or seven laser beams with identical linear
polarizations are employed to obtain 
interference patterns with decagonal or tetradecagonal rotational symmetry, 
respectively (see Fig.\ \ref{fig:pot}). For vanishing tilt angle 
$\theta \rightarrow 0$ of the beams, the potential in the $xy$ plane 
is \cite{schmiedeberg07,gorkhali05}:
\begin{equation}
\label{eq:pot}
V(\mathbf{r})=-\frac{V_0}{N^2}\sum_{j=0}^{N-1} \sum_{k=0}^{N-1} 
{\cos\left[\left(\mathbf{G}_j-\mathbf{G}_k\right) \cdot \mathbf{r} +
\phi_j-\phi_k\right]},
\end{equation}
where $N$ is the number of beams, $\phi_j$ are the phases of the 
laser-light waves, and $\mathbf{G}_j$ their wave-vectors projected onto the 
$xy$ plane [cf.\ Fig.\ \ref{fig:pot}(b) and (f)]. The prefactor is chosen 
such that $-V_0$ gives the minimum value of the potential. We usually 
set $\phi_j=0$ for all $j$ corresponding to a laser field without any 
phononic or phasonic displacements
and a center of exact 10- or 14-fold rotational symmetry at 
$\mathbf{r} = \mathbf{0}$.
In addition, for the decagonal potential, 
we explicitly analyze the effect of phasonic distortions on the 
Archimedean-like tiling in the last section. In this case, the phases $\phi_j$ 
are used to specify the phononic displacement field $\mathbf{u}
(\mathbf{r},t ) = [u_x (\mathbf{r},t ), u_y(\mathbf{r},t)]$ and the 
phasonic field $\mathbf{w}(\mathbf{r},t) = [w_x(\mathbf{r},t),
w_y(\mathbf{r},t)]$ following the convention of Ref.\ \cite{levine85a,levine85b}, 
\begin{equation}
\label{eq:phason}
\phi_j\left(\mathbf{r},t\right)=\mathbf{u}\left(\mathbf{r},t\right)\cdot 
\mathbf{G}_j + \mathbf{w}\left(\mathbf{r},t\right) \cdot 
\mathbf{G}_{3j\textnormal{\ mod\  }5}.
\end{equation}
In section \ref{sec:phasons} we will show how a uniform phasonic displacement
$\mathbf{w}(\mathbf{r},t)$, which increases linearly in time, or how a 
static $\mathbf{w} (\mathbf{r},t)$ with a constant gradient can be used 
to align the rows of the Archimedean-like tiling phases along
specific directions of the underlying substrate potential.

\section{Methods} 
\label{sec:methods}

In the following, we shortly introduce our methods, which we employed to
analyze the results from experiments and simulations.

\subsection{Experiment} 

The colloidal system we used is an aqueous suspension of highly charged 
sulphate-terminated polystyrene particles with a radius of 
$R=1.45\ \mu\textnormal{m}$ from Interfacial Dynamics Corporation with an 
average polydispersity below 4\%. The pair interaction of particles can 
be described by a screened Coulomb potential as introduced below in
Eq. (\ref{eq:pairpot}). As sample cell we used a 
silica glass cuvette with $200\ \mu\textnormal{m}$ spacing. 
Before starting the measurement, it was connected to a deionization circuit to reduce the salt concentration of the suspension. This circuit also included an electrical conductivity probe (typical ionic conductivities below $0.08\ \mu\textnormal{S/cm}$), a vessel of ion exchange resin, and a peristaltic pump. After deionization, the suspension was inserted into the cell which was then sealed. Before applying the quasiperiodic potential, the density of the particles was increased in the field of view. This is achieved through thermophoretic and optical forces using vertically incident fiber coupled infrared laser with a wavelength of $\lambda=1070\ \textnormal{nm}$ and maximum output power $P_{\textnormal{max}}=5\ \textnormal{W}$ and an argon ion laser ($\lambda=514\ \textnormal{nm}$, $P_{\textnormal{max}}=7\ \textnormal{W}$). A circular monolayer was formed on the lower surface of the sample cell with a diameter of about $500\ \mu\textnormal{m}$. The vertical fluctuations of the particles are also suppressed by the argon ion beam, therefore the system can be considered as two-dimensional. The monolayer spontaneously adapts the hexagonal packing. After reaching the desired particle density, the infrared laser was turned off. The high particle density was kept constant using an optical tweezer ($\lambda=488\ \textnormal{nm}$, $P_{\textnormal{max}}=7\ \textnormal{W}$) scanned around the central region of the system, i.e., forming a circular corral.
The quasiperiodic potential was created by interference of five or 
seven linearly polarized beams of a Nd:YVO4 laser
($\lambda=532\ \textnormal{nm}$, $P_{\textnormal{max}}=18\ \textnormal{W}$). 
Due to optical gradient forces, this interference pattern acts as an 
external potential on the particles. 
The depth $V_0$ of the singular deepest potential well 
(see Sec.\ \ref{sec:model}) scales linearly with the laser power.
$V_0$ was determined by calibrating the laser intensity for three laser 
beams, i.e., when they create a triangular potential. The length scale 
$a_V$ was tuned by changing the angle of incidence of the laser beams
with respect to the sample plane.
The different colloidal phases were identified with the help of 
a Delaunay triangulation. It creates nearest-neighbor bonds between
vertices that were defined by the maxima of the colloidal density 
distribution.
For Archimedean-like tiling structures the bond length is a bimodal 
distribution with the ratio of the peak positions close to $\sqrt{2}$. 
Bonds longer than $1.2a_S$ 
were removed from the triangulation which resulted in the 
characteristic square-triangle tiling.

\subsection{Simulations} 

The interaction between the colloids is 
modeled by the pair potential 
\begin{equation}
\phi(d)=\frac{[Z^{*}e\exp(\kappa R)]^2}{4\pi\epsilon_{0}
\epsilon_{r}(1+\kappa R)^2} \frac{\exp(-\kappa d)}{d}
\label{eq:pairpot}
\end{equation}
of the Derjaguin-Landau-Verwey-Overbeek theory \cite{Derjaguin41,Verwey48}, 
where $d$ is the distance between two colloids and $\kappa$ the inverse 
Debye screening length. The prefactor also depends on the radius $R$ of a 
colloid, its effective surface charge $Z^{*}$, and the dielectric constant 
of water $\epsilon_{r}$. 

We quantify the particle density by the spacing $a_{S}$ of the particles
when placed on
an ideal triangular lattice. The occurrence of Archimedean-like tilings 
mainly depends on the density of the system. All other parameters can be 
varied over huge ranges. The main limitation to the choice of parameters 
is that there has to be a competition between the 
colloidal interactions and the interaction with the substrate. 
Therefore, we usually 
choose the parameters such that the density used in the simulations
is slightly above the density where the triangular to liquid phase 
transition occurs.
For example, the parameter set for the decagonal substrate employed already 
in \cite{schmiedeberg08} is suitable to observe Archi\-medean-like tilings, 
i.e., $R=1.2\ \mu\textnormal{m}$, $Z^{*}=1000$, $\epsilon_{r}=78$, $T=300\ 
\textnormal{K}$, $a_V=5.0\ \mu\textnormal{m}$, and $\kappa^{-1}=0.25\ \mu
\textnormal{m}$. Whereas a parameter set closer to the experimental values 
$R=1.45\ \mu\textnormal{m}$, $Z^{*}=400000$, $\epsilon_{r}=78$, 
$T=300\ \textnormal{K}$, $a_V=6.5\ \mu\textnormal{m}$, and 
$\kappa^{-1}=0.2\ \mu\textnormal{m}$ (cf.\ \cite{mikhael09}) was employed
to simulate colloidal ordering in the tetradecagonal laser field. 
We usually performed Monte-Carlo simulations 
using the Metropolis algorithm \cite{Metropolis53}.
Periodic boundary conditions were implemented for simulation boxes
that contained about 500 to 1400 colloids. For comparison or to study 
dynamical properties, we also employed Brownian dynamics simulations as 
specified in \cite{schmiedeberg07}.

When domains with Archimedean-like tiling structures 
occur, 
their rows are aligned along the respective five or seven
equivalent directions of the laser potential.
Usually within one simulation box, 
one finds many domains with different orientations. There is a tendency that small domains anneal out in time, i.e., adjacent domains grow slowly within the simulation process. The structures shown 
in Fig.\ \ref{fig:AT-sim} are about the largest domains we achieved by
pure Monte-Carlo simulations. However, we discovered different methods to 
support the formation of larger domains. 
While in Monte-Carlo simulations the direction of a domain may change 
quite easily, such domain flips are rare in Brownian dynamics simulations. 
Therefore, Brownian dynamics simulations can help to grow large uniform 
domains. It is also possible to predefine a preferred direction. For example, 
by using small simulation boxes where the potential is discontinuous at 
the boundaries, the rows of an Archimedean-like tiling start to form along 
the boundaries. Furthermore, phasonic distortions 
support the growth of Archi\-medean-like tilings along 
specific directions as discussed in Sec.\ \ref{sec:phasons}.

\section{Archimedean-like tilings} 
\label{sec:structure}

\begin{figure}
\begin{center}
  \includegraphics[width=0.65\linewidth]{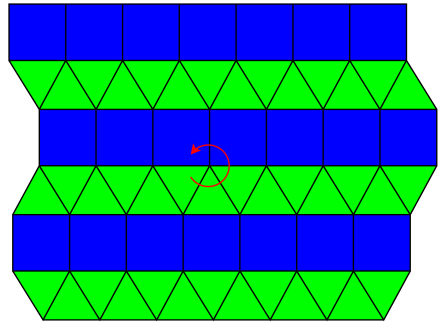}
\end{center}
  \caption{Archimedean tiling of type $(3^3\cdot4^2)$.}
  \label{fig:at} 
\end{figure}

The structures we observe 
in the two-dimensional quasicrystalline potentials at certain particle
densities
consist of rows of squares and rows of triangles and therefore are similar 
to the Archimedean tiling shown in Figure \ref{fig:at}. In general, 
Archimedean tilings are formed by regular polygons and are characterized 
by the number of edges of the polygons that meet in each vertex 
\cite{gruenbaum}. The tiling shown in Figure \ref{fig:at} is called 
$(3^3\cdot4^2)$-Archimedean tiling, because going around a vertex, one first 
finds three regular triangles and then two squares (see red arrow in 
Fig.\ \ref{fig:at}). Very recently similar Archimedean tilings were also observed for 
binary mixtures of nanoparticles 
situated between a cubic and a dodecagonal 
phase \cite{talapin}. Furthermore, an Archimedean tiling of type $(3^2\cdot 4\cdot 3\cdot 4)$
occured in simulations of monoatomic systems on periodic substrates 
\cite{patrykiejewa,patrykiejewb,patrykiejewc,patrykiejewd}.

\subsection{On decagonal substrates} 
\label{sec:stru10}

The Archimedean-like tiling on a decagonal substrate differs from the 
perfect $(3^3\cdot 4^2)$-Archimedean tiling by 
the frequent appearance of
double rows of triangles [see Figs.\ \ref{fig:AT-sim}(a,b)]. 
From the simulation data
we were able to identify the  sequence over a range of 
$-6a_V\leq y \leq 100a_V$ that correspond to 77 single or double rows of
triangles. Within this range the 
observed sequence is exactly the sequence of the Fibonacci chain, if one 
identifies a double row with $L$ and a single row with $S$ [cf.\ Fig.\ 
\ref{fig:AT-sim}(b)]. A Fibonacci chain is obtained by starting with 
element $L$ and repeatedly applying the iteration rules $L\rightarrow LS$ 
and $S\rightarrow L$.
For example, the first iterations are $L$, $LS$, $LSL$, 
$LSLLS$, $LSLLSLSL$, etc.. 

Note, the squares and triangles of the Archimedean-like tiling are not
perfect regular polygons. A row of triangles
has almost the same height as a row of squares. 
In the simulations, we find
the ratio of the mean row height $l_y$ divided by the
mean  particle spacing $l_x$ in the direction along the rows to be
$l_y/l_x\approx 0.95$, i.e., the triangles are
stretched in the direction perpendicular to the rows while the squares
are stretched in the direction along the rows. We will calculate the ratio
$l_y/l_x$ 
in subsection \ref{sec:density}.

At a first glance, 
the Archimedean-like tiling structure seems to be periodic along the rows 
of triangles.
However, some aperiodic modulations along the rows can be observed. For 
example, in the structures shown in Figs.\ \ref{fig:AT-sim}(a,b) only very 
few bonds are perfectly horizontal. Whereas most of the bonds between two 
neighboring rows of triangles are almost horizontal, the direction of all 
other bonds deviate from the horizontal direction in an aperiodic way. 
Furthermore, there is a modulation of the bond lengths along the direction of 
the rows. 
As a result, some rows of squares exist along a direction that is rotated 
by $2\pi/5$ from the main horizontal row direction.

\subsection{On tetradecagonal substrates} 

\begin{figure}
\begin{center}
  \includegraphics[width=0.95\linewidth]{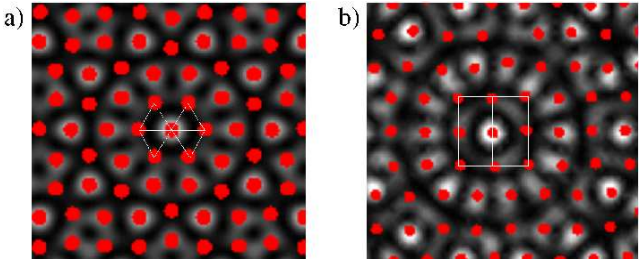}
\end{center}
  \caption{Colloidal order close to symmetry centers in simulations: 
(a) On decagonal substrates double rows of triangles occur at symmetry 
centers, (b) in tetradecagonal laser fields double rows of squares are 
found.}
  \label{fig:ATstars} 
\end{figure}

The Archimedean-like tiling phase on a tetradecagonal substrate consists of 
large periodic regions that correspond to a perfect Archimedean tiling of 
type $(3^3\cdot4^2)$. The periodicity is only interrupted by a few double 
rows of squares. Fig.\ \ref{fig:ATstars}(b) shows the position of the colloids 
on the substrate. 
Obviously, a double row of squares occurs at a local symmetry center of the 
laser field. This bright spot corresponds to a deep potential minimum that 
is surrounded by 14 shallower potential wells. It therefore constitutes
the center of a region with 14-fold rotational symmetry. 
Since in a tetradecagonal laser field local symmetry centers are rare 
\cite{mikhael09}, only few double rows of squares exist in the 
Archimedean-like tiling structure. Fig.\ \ref{fig:ATstars}(a) demonstrates
that for the decagonal substrate double rows of triangles occur at symmetry 
centers. Since there are much more symmetry centers in a decagonal laser 
field \cite{mikhael09}, much more double rows 
exist in the 
Archimedean-like tiling phase of a decagonal substrate compared to a 
tetradecagonal laser field.

\subsection{Densities with Archimedean-like tilings} 
\label{sec:density}

\begin{figure}
\begin{center}
  \includegraphics[width=0.7\linewidth]{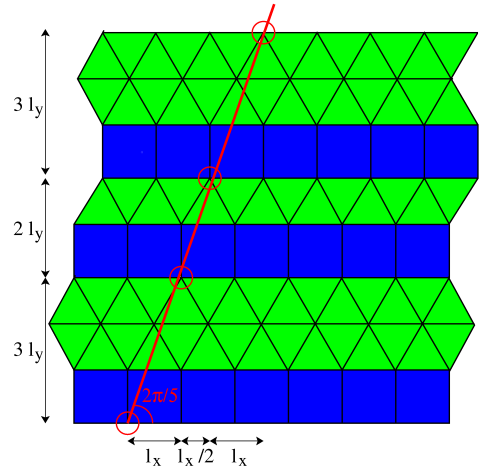}
\end{center}
  \caption{We calculate the mean bond length in horizontal direction 
$l_x$ in a way that many colloids lie along the $2\pi/5$-direction 
(circled red). The mean row hight $l_y$ is chosen to fit one of the two 
length scales of the substrate in $y$-direction.}
  \label{fig:AT-dichte} 
\end{figure}

In the following, we want to estimate
the densities where Archimedean-like 
tiling structures occur. At a suitable density, the rows of colloids 
should be oriented
along lines with many minima of the potential landscape. In the 
decagonal laser field, we can determine the distances between such lines 
by averaging the potential $V(\mathbf{r})$ along the $x$-direction, i.e.,
\begin{eqnarray}
\label{eq:proj}
\langle V\rangle_x=&&\frac{V_0}{25}\left[-5-2\cos\left(2\pi y/\lambda_1\right)
-2\cos\left(2\pi y/\lambda_2\right)\right],\\
\nonumber
\textnormal{with\ }&&\lambda_1=\sqrt{\frac{5-\sqrt{5}}{10}}\ a_V\\
\nonumber
\textnormal{and\ }&&\lambda_2=\sqrt{\frac{5+\sqrt{5}}{10}}\ a_V.
\end{eqnarray}
We now calculate the density of an Archimedean-like tiling whose mean 
row height $l_y$ fits to one of the lengths $\lambda_1$ or $\lambda_2$. 

The simulation results show that the mean bond length $l_x$ in the direction 
of the rows differs from the row height $l_y$. Therefore,
in the following, we aim at calculating the ratio $l_y/l_x$.
So far we have determined the row height $l_y$ of an Archimedean-like 
tiling that fits to the decagonal substrate. To estimate the bond length 
$l_x$, we argue as follows. The rows are oriented along a symmetry 
direction of the decagonal potential. On this direction many potential 
minima are situated that are occupied by the particles. However, an 
equivalent symmetry direction is found at an angle of $2\pi/5$ with the 
rows. It should also be occupied by as much particles as possible.
Figure\ \ref{fig:AT-dichte} demonstrates how this direction can be 
fitted to the Archimedean-like tiling. Colloids close to the red line
are marked by red circles. In case of a single row of triangles, to move
from one particle to another close to the red line, one steps $l_x/2$ in 
$x$-direction and $2l_y$ in $y$-direction.  In case of a double row of 
triangles, the step lenghts are $l_x$ in $x$-direction and $3l_y$ in 
$y$-direction. The sequence of single and double rows corresponds to 
the sequence of the Fibonacci chain. Therefore, there are $\tau$ times 
as many double rows as single rows of triangles, where 
$\tau=\left(1+\sqrt{5}\right)/2\approx 1.618$ is the number of the 
golden mean. As a consequence, the orientation of the red line is
determined by
\begin{equation}
\label{eq:tan}
\tan\left(2\pi/5\right)=\frac{\left(2+3\tau\right)l_y}{\left(\frac{1}{2}
+\tau\right)l_x},
\end{equation}
i.e., we find the ratio $l_y/l_x\approx 0.951$ which agrees remarkably well with the 
value determined from simulations in subsection \ref{sec:stru10}.

To calculate the mean particle spacing $a_S$, we compare the number 
density in the Archimedean-like 
tiling,
\begin{equation}
\label{eq:dichte}
\rho=\frac{1}{l_xl_y} \enspace,
\end{equation}
with the number density of a triangular lattice $\rho=2/\left(\sqrt{3}
a_S^2\right)$. By using $l_x$ from Eq.\ (\ref{eq:tan}), 
we find for the mean particle spacing of the Archimedean-like tiling,
\begin{equation}
a_S=\sqrt{\frac{2}{\sqrt{3}}\frac{2+3\tau}{\frac{1}{2}+\tau}\frac{1}{\tan
\left(2\pi/5\right)}}\,l_y=\left[\frac{8}{15}\left(5-\sqrt{5}\right)
\right]^{\frac{1}{4}}l_y,
\end{equation}
where $l_y$
has to be one of the length scales $\lambda_{1,2}$ introduced in 
Eq.\ (\ref{eq:proj}). Therefore,
Archimedean-like tilings on the decagonal substrate occur for 
\begin{eqnarray}
\nonumber
a_S/a_V&=&2\left[\frac{1}{75}\left(5-2\sqrt{5}\right)\right]^{\frac{1}{4}}
\approx 0.579\\
\label{eq:dens-deca}
\textnormal{\ or\ } a_S/a_V&=&\left[\frac{8}{75}\left(5+\sqrt{5}\right)
\right]^{\frac{1}{4}}\approx 0.937.
\end{eqnarray}
Indeed, in experiments and in simulations, we find the Archimedean-like 
tiling structures for densities close to these theoretical values (see, 
e.g., phase diagram in Fig.\ \ref{fig:pd-at}).

\begin{figure}
\begin{center}
  \includegraphics[width=0.8\linewidth]{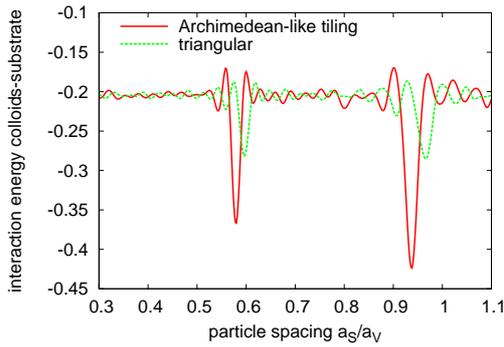}
\end{center}
  \caption{Average potential energy of a colloid in the decagonal potential 
calculated for a triangular and for an Archimedean-like tiling structure 
as a function of density, which is given by the particle spacing $a_S/a_V$. To determine the average potential energy, we constructed a perfect triangular or Archimedean-like tiling structure with the corresponding particle spacing and calculated the average potential depth at the positions of the colloids of the structure.}
  \label{fig:at5-en}
\end{figure}

In Fig.\ \ref{fig:at5-en} we 
show how the average potential energy of a colloid varies with density
when it is part of either a triangular lattice or an Archimedean-like 
tiling structure under the influence of the decagonal substrate potential.
We assumed that the Archimedean-like tiling structure is stretched 
perpendicular to the direction of the rows such that Eq.\ (\ref{eq:tan})
holds. From Fig.\ \ref{fig:at5-en}
we can clearly identify
two densities where the Archimedean-like tiling fits to the 
substrate very well.
Most of the colloids are located in minima of the potential and therefore 
the average potential energy exhibits two sharp minima.
These densities correspond to the 
values calculated in Eq.\ (\ref{eq:dens-deca}). Note, we also determined the average potential energy for other Archimedean tilings, but did not find any other structure that shows pronounced minima. For example, an Archimedean tiling of type $(3^2\cdot 4\cdot 3\cdot 4)$ has exactly the same particle interaction energy as the tiling of type $(3^3\cdot 4^2)$. However, by comparing the average potential energies, we find that the substrate favors the row structure corresponding to type $(3^3\cdot 4^2)$.

\begin{figure}
\begin{center}
  \includegraphics[width=0.8\linewidth]{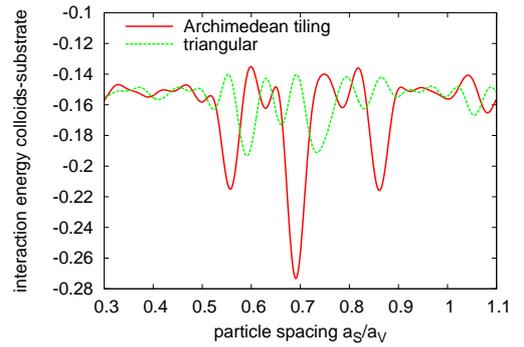}
\end{center}
  \caption{Average potential energy of a colloid in the tetradecagonal 
potential calculated for a triangular and for an Archimedean-like tiling 
structure as a function of density, which is given by the particle spacing 
$a_S/a_V$.}
  \label{fig:at7-en} 
\end{figure}

To determine the densities, where Archimedean-like tilings occur 
under the influence of
the tetradecagonal substrate potential, we calculate the average 
potential energy of a colloid within an Archimedean tiling structure 
as a function of
the particle spacing (see Fig.\ \ref{fig:at7-en}). 
Again, we
find 
pronounced minima of the potential energy (although not as sharp as for
decagonal potentials)
at
$a_S/a_V=0.557$, $a_S/a_V=0.691$, and $a_S/a_V=0.861$. At these
values we were indeed able to observe Archimedean-like tilings in the 
simulations.

\begin{figure}
\begin{center}
  \includegraphics[width=0.7\linewidth]{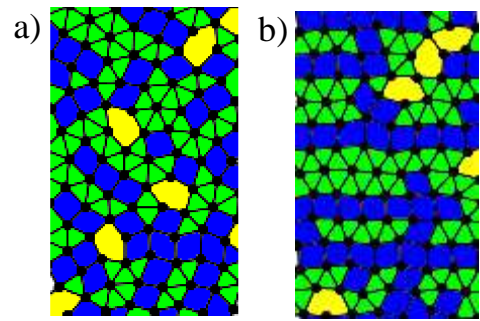}
\end{center}
  \caption{Brownian dynamics simulation without confining boundaries
started at an initial density
that corresponds to $a_S/a_V=0.565$.
(a) Delaunay construction (as in Fig.\ \ref{fig:AT-sim}) displaying the 
colloidal configuration when the simulation starts ($t=0$). After time 
$t=0.05 \gamma a_V^2 /k_BT$, where $\gamma$ is the friction coefficient 
of a colloid,
an Archimedean-like tiling structure can be observed which remains almost 
unchanged up to the end of our simulation at time 
$t=0.25 \gamma a_V^2/k_BT$ shown in (b). Note, while the Archimedean-like tilings shown in Fig. \ref{fig:AT-sim} are equilibrium phases that occur in systems with densities fixed at special values, the structure in (b) is only metastable because it will slowly dissolve when the density decreases due to the missing boundaries.}
  \label{fig:ATdyn} 
\end{figure}

We just argued that Archimedean-like tilings should only occur at 
specific densities. Correspondingly, when using Monte-Carlo simulations to find the equilibrium phases, we only observe them close to the calculated densities, e.g., in a 
narrow density region in the phase diagram of Fig.\ \ref{fig:pd-at}.
Whereas in the simulations the density is a constant quantity, it 
can to a certain degree change its value in experiments; the 
experimental system is large enough that the density can become
non-uniform and colloids can leave the boundary box. 
Indeed, when in the experiment we bring the density close to the ones 
predicted by Eq.\ (\ref{eq:dens-deca}), we observe that the colloids
self-adjust their density so that the Archimedean-like tiling forms. Figure\ \ref{fig:at5-en} explains this observation since the energy is 
lowered by forming an Archimedean-like tiling, at least relative
to the triangular phase.
To further investigate the idea of 
self-adjusting the colloidal density, we performed the following Brownian 
dynamics simulations: We confine the colloids at a density given by 
$a_S/a_V=0.565$ that is larger than the one suitable for Archimedean-like 
tilings ($a_S/a_V=0.58$). We then remove the confining boundaries so that
the colloids are able to float over the substrate potential and the system
expands. Indeed, we find that particles adjust to the potential by 
forming the characteristic rows of squares and triangles of an 
Archimedean-like tiling (see Fig.\ \ref{fig:ATdyn}). Of course, 
this structure is not the thermodynamic ground state in such an open 
system. However, it remains present during the duration of the 
simulation which is the time a free particle would need to diffuse 
a distance $a_V$. Only at the edges of an Archimedean-like tiling domain
does the structure start to dissolve. This demonstrates while Archimedean-like tilings are equilibrium phases at special densities only, they also occur as metastable structures in systems with open boundaries.

\section{Phasonic Distortions} 
\label{sec:phasons}

In this section we 
demonstrate, how the ordering of colloids in the Archimedean-like tiling 
that forms 
in the decagonal laser field is affected by 
manipulating the phasonic degrees of freedom as defined in 
Eq.\ (\ref{eq:phason}). The consequences 
of a constant phasonic displacement on a quasicrystalline phase with 
decagonal symmetry were presented in \cite{schmiedeberg08}.

\subsection{Inducing Phasonic Flips} 

\begin{figure}
\begin{center}
  \includegraphics[width=0.75\linewidth]{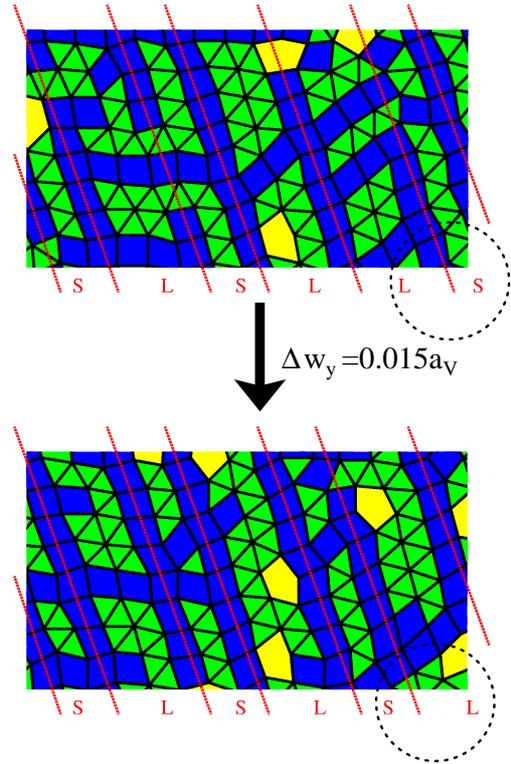}
\end{center}
  \caption{Inducing a phasonic flip: Delaunay construction (as in Fig.\ \ref{fig:AT-sim}) for the Archimedean-like 
tiling phase ($a_S/a_V=0.577$, $V_0/(k_BT)=20$) obtained by Monte-Carlo 
simulations for $w_y=0$ (top) and $w_y=0.015 a_V$ (bottom). The flip in the 
Fibonacci chain defined by the single and double rows of triangles 
is marked by dotted circles.}
  \label{fig:phasonen-AT} 
\end{figure}

We first 
study how a constant phasonic displacement influences
the Archimedean-like tiling. Figure \ref{fig:phasonen-AT} shows patterns 
of the Archimedean-like tiling in the decagonal laser field for different 
phasonic 
components $w_y=0$ and $w_y = 0.015a_V$.
Note, a whole row of triangles and a row of squares have interchanged 
their position. This corresponds to a phasonic flip in the Fibonacci chain
defined by the sequence of long and short distances between the square rows.

Figure \ref{fig:pot}~(d) shows very clearly sets of parallel lines of 
low intensity in the decagonal laser field that are oriented along its
five equivalent directions. One can choose phasonic displacements such 
that one set of dark parallel lines does not change. So if the square 
and triangular rows are parallel to this set of lines, the whole 
Archimedean-like tiling is not affected by the phasonic displacement. 
For example, a phasonic displacement $w_x$ does not influence the square and 
triangular rows when they are oriented along the $x$ direction. As we show 
in the next subsection, a  phasonic drift in $w_x$ may even help to 
stabilize the Archimedean-like tiling structure.

\subsection{Stabilization by Phasonic Drifts and Gradients} 

\begin{figure}
\begin{center}
  \includegraphics[width=0.95\linewidth]{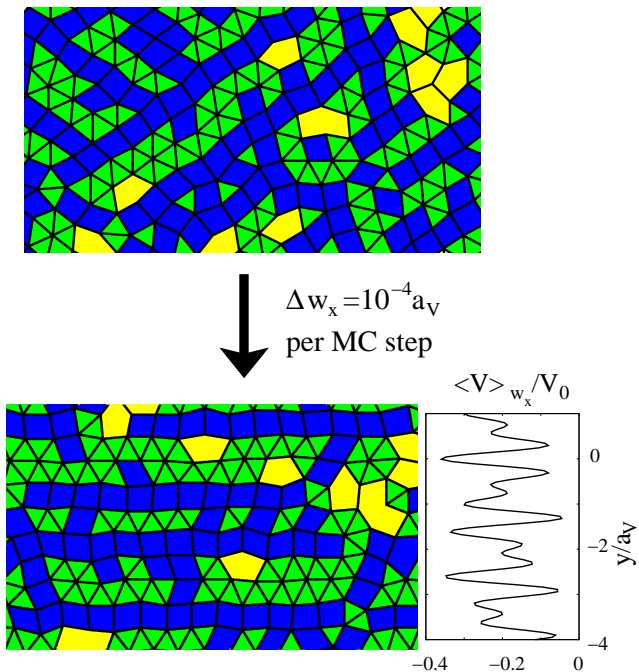}
\end{center}
  \caption{Stabilization by a phasonic drift: 
Archimedean-like tiling phase (same parameters as in Fig.\ 
\ref{fig:phasonen-AT}) obtained 
by Monte-Carlo simulations without phasonic displacement (top) and in 
a potential with a phasonic drift velocity of $\Delta w_x=10^4a_V$ per 
Monte-Carlo step (bottom). The small graph on the right-hand side shows 
the decagonal potential averaged over $w_x$ as a function of $y$.}
  \label{fig:phasonen-stab} 
\end{figure}

A steadily growing $w_x$ in time orients the Archimedean-like tiling along the $x$-direction. 
This is demonstrated in Fig.\ \ref{fig:phasonen-stab} 
for a phasonic drift velocity
$\Delta w_x=10^{-4}a_V$ per unit time.
Domains that are not oriented along the $x$-direction are reduced 
since a nonzero $\Delta w_x$ constantly induces phasonic flips in the
rows of squares and triangles of the Archimedean-like tiling unless they
are oriented along the $x$-direction.
Another way of explaining the consequences of the phasonic drift velocity
$\Delta w_x$ goes as follows. One can demonstrate that a steadily growing 
phasonic displacement rearranges the decagonal potential landscape such that 
existing potential wells disappear and reappear at other locations.
Now consider a colloid in the decagonal laser field that experiences a 
sufficiently fast phasonic drift in $w_x$-direction, i.e., the relaxation
time of the colloid within a potential well is much larger than the
time scale on which the potential well vanishes due to the phasonic drift.
Then one can introduce an effective potential $\langle V\rangle_{w_x}(y)$
for the colloids that is the decagonal potential $V(x,y)$ averaged over 
$w_x$. It is illustrated at the right hand side of 
Fig.\  \ref{fig:phasonen-stab}. From Eq. (\ref{eq:pot}) it is clear that 
$\langle V\rangle_{w_x}(y)$ does not depend on $x$ or equals $V(x,y)$
averaged over the spatial coordinate $x$ already introduced in 
Eq.\ (\ref{eq:proj}), i.e., 
\begin{equation}
\langle V\rangle_{w_x}(y)=\langle V\rangle_{x}(y).
\end{equation}
Clearly, such an effective potential can only support domains of 
Archimedean-like tilings oriented along the $x$-direction. 

\begin{figure}
\begin{center}
  \includegraphics[width=0.7\linewidth]{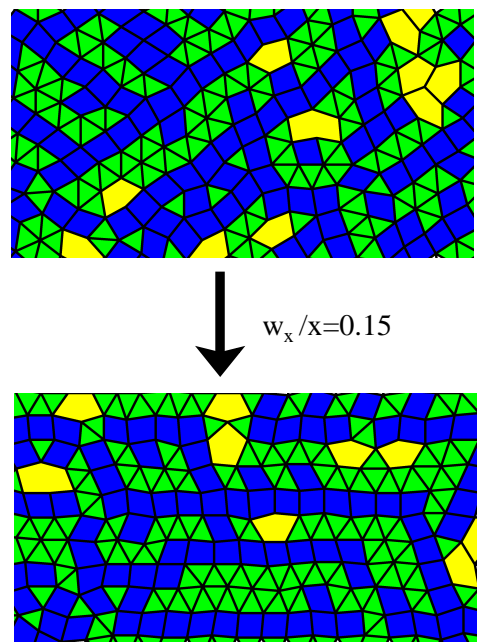}
\end{center}
  \caption{Stabilization by a phasonic gradient: 
Archimedean-like tiling phase (same parameters as in Fig.\ 
\ref{fig:phasonen-AT}) obtained by Monte-Carlo simulations
without phasonic displacement (top) and in a potential with a 
phasonic gradient with $w_x/x=0.15$ (bottom).}
  \label{fig:phasonen-grad} 
\end{figure}

\begin{figure}
\begin{center}
  \includegraphics[width=0.95\linewidth]{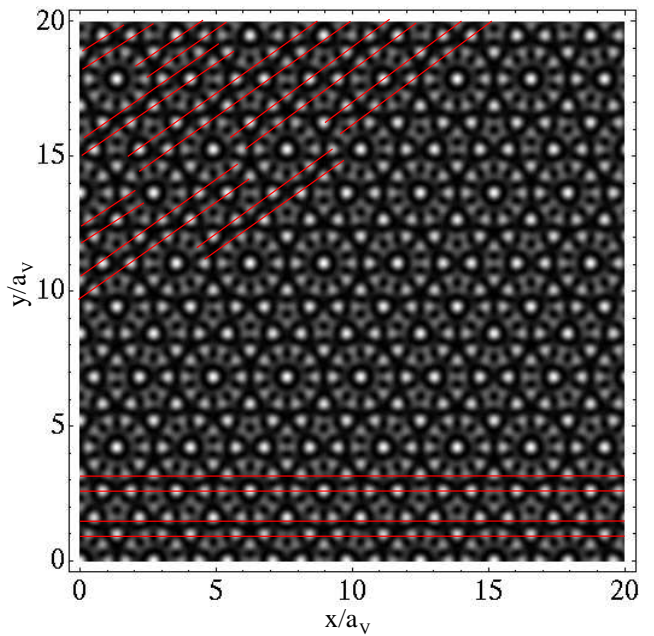}
\end{center}
  \caption{Interference pattern of the decagonal laser field calculated
from Eq. (\ref{eq:pot}) when the phase $\phi_0$ attached to wave vector 
$\mathbf{G}_0$ varies with a constant gradient, $\phi_0=0.2x/a_V$.
Via Eq.\ (\ref{eq:phason}) one can show that it creates a constant gradient 
in the $x$ components of the phononic and phasonic displacements. As a result,
the lines of low intensity are infinitly extended in $x$-direction
whereas in all other directions they have a finite length (see, e.g., 
the dark stripes framed by the red lines). Similar jags are studied in \cite{freedman07a,freedman07b}.
}
  \label{fig:pot-phason} 
\end{figure}

Another possibility of orienting domains of Archimedean-like tilings
preferentially along the $x$-direction is to introduce a phasonic 
displacement field with a constant gradient in $x$-direction.
Figure\ \ref{fig:pot-phason} shows the resulting decagonal interference
pattern. The gradient in $w_x$ destroys all continuous lines of low 
intensity in the potential landscape, except those in $x$-direction, as
indicated by the red lines.
Therefore, the Archimedean-like tiling is mainly oriented along the 
direction of the phasonic gradient, as illustrated in Fig. 
\ref{fig:phasonen-grad}.

In the experimental interference pattern of Fig. \ref{fig:pot}c) some of
the lines of low intensity just end in the middle of the pattern. This 
suggests that phasonic distortions are present, as we will now demonstrate.
Suppose that the five laser beams in Fig. \ref{fig:pot}a) are not 
perfectly arranged around the vertical. For example, if the tilt angle 
$\theta_0$ of beam 0 is larger than the other angles, the projected wave 
vector becomes $\mathbf{G}'_{0}=\mathbf{G}_{0}+\Delta\mathbf{G}_{0}$
with $\left|\mathbf{G}'_{0}\right| > \left|\mathbf{G}_{i}\right| 
(i=0,1 ... 4)$. From Eq.\ (\ref{eq:pot}) one finds that the deviation
$\Delta\mathbf{G}_{0}$ can be interpreted as a constant gradient in the 
phase $\phi_0$ attached to $\mathbf{G}_{0}$, 
$\phi_0=\Delta\mathbf{G}_{0}\cdot\mathbf{r}$. Using Eq.\ (\ref{eq:phason}),
this phase $\phi_0$ can be decomposed into $x$ components of the phononic 
and phasonic displacement fields with constant gradients. 
Therefore, laser 
beams that are not perfectly adjusted relative to the vertical lead to a 
substrate potential with phononic and phasonic distortions. Furthermore, such distortions can also be controlled intentionally by
displacing one of the laser beams radially from its position
on an ideal pentagon. This way one can create a preferred
orientation for the Archimedean-like tiling, as discussed
above.

In summary, phasonic drifts or distortions can be employed to grow large domains of an Archimedean-like tiling structure with a preferred orientation. These domains are stable when the phasonic drift is stopped or the phasonic distortion is removed.

\section{Outlook and Conclusions}    \label{sec:conclusions}

\begin{figure}
\begin{center}
  \includegraphics[width=0.85\linewidth]{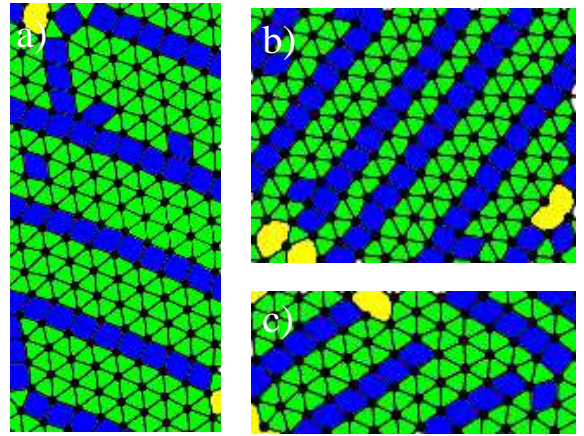}
\end{center}
  \caption{Delaunay  construction (as in Fig.\ \ref{fig:AT-sim}) of 
simulation results that consist of rows of triangles and rows of squares 
in laser fields (of strength $V_0/(k_BT)=50$) created by (a) eight beams 
($a_S/a_V=0.8$), (b) ten beams ($a_S/a_V=0.7$), and (c) eleven beams 
($a_S/a_V=0.74$).}
  \label{fig:ATn} 
\end{figure}

In this article we have studied the structure of Archimedean-like tiling 
phases that form in decagonal and tetradecagonal laser fields at certain
densities. We have also demonstrated how the domain size of such tilings 
is influenced by phasonic distortions. In particular, if the interfering 
laser beams are not adjusted properly, phasonic displacement fields with
a constant gradient arise. Interestingly, simulations using substrate
potentials with further quasicrystalline symmetries also show similar 
phases. For example, in an interference pattern of eight beams we find 
rows of squares and triple or four-fold rows of triangles 
[Fig.\ \ref{fig:ATn}(a)]. A laser field created by ten beams favors 
exactly the same structure as the decagonal potential obtained by five 
laser beams [Fig.\ \ref{fig:ATn}(b)]. For the case of eleven beams, 
we also observe the characteristic rows of triangles and squares
[Fig.\ \ref{fig:ATn}(c)]. However, we are not able to analyze this 
ordering in more detail with respect to the order perpendicular to
the rows, since the domains are too small. It is an interesting question 
for future research to understand, why such Archimedean-like tilings
form at specific densities on substrates with different quasicrystalline 
symmetries. Depending on the specific rotational symmetry, all these 
structures consist of single and double (or even triple or four-fold) rows 
of triangles or squares that ultimately form a quasiperiodic order in one 
direction.

While on the decagonal substrate the rows in the Ar\-chi\-me\-dean-like 
tiling  form a Fibonacci chain, they show domains with periodic order on
the tetradecagonal substrate interrupted by double rows of squares.
We find that these double rows, as the double rows of triangles on the
decagonal substrate, occur at symmetry centers of the substrate
potential. As discussed in \cite{mikhael09}, the number of the symmetry 
centers is by a factor of 100 larger in the decagonal compared to
the tetradecagonal substrate. Therefore, they enforce the formation of
a Fibonacci chain in the first case, whereas in the tetradecagonal
laser field larger periodic domains form due to the much smaller
number of symmetry centers.

Amongst other phases (see Ref. \cite{schmiedeberg08}), Archimedean-like 
tilings appear as a compromise between structures favored either by the 
particle interaction or by the substrate potential. There is some evidence 
that they occur in experiments that study how atoms order on atomic surfaces 
with decagonal symmetry \cite{mikhael08}. For the specific
example of the Archimedean-like tiling, we have presented here Brownian
dynamics simulations to demonstrate how the colloidal adsorbate reacts on 
phasonic distortions in the substrate. This rises the question if phasonic 
excitations are detectable in atomic substrates. The Archimedean-like
tiling as an adsorbate phase
might be a good candidate to visualize these excitations in the 
underlying substrate since resulting rearrangements in the rows of 
triangles and squares should be visible.

\begin{acknowledgement}
We would like to thank R. Lifshitz and H.-R. Trebin for helpful discussions. 
We acknowledge financial support from the Deutsche Forschungsgemeinschaft 
under Grant No.\ RO 924/5-1 and BE 1788/5.
\end{acknowledgement}

\end{document}